\documentclass[conference]{IEEEtran}
\makeatletter
\def\ps@headings{%
\def\@oddhead{\mbox{}\scriptsize\rightmark \hfil \thepage}%
\def\@evenhead{\scriptsize\thepage \hfil \leftmark\mbox{}}%
\def\@oddfoot{}%
\def\@evenfoot{}}
\makeatother
\usepackage{epsfig}
\usepackage{amsfonts}
\usepackage{amssymb}
\newcommand {\mymarginpar}[1]{\marginpar{#1}}
\renewcommand {\marginpar}[1]{}

\newtheorem{axiom}{Assumption}
\newtheorem{definition}{Definition}
\newtheorem{property}[definition]{Property}
\newtheorem{proposition}[definition]{Proposition}
\newtheorem{lemma}[definition]{Lemma}
\newtheorem{theorem}[definition]{Theorem}
\newtheorem{corollary}[definition]{Corollary}

\newtheorem{remark}[definition]{Remark}

\newcommand {\bsec}[2]{\section
                       {#1}
                       \label{sec:#2}}

\newcommand {\rsec}[1]{Section \ref{sec:#1}}

\newcommand {\benum} {\begin{enumerate}}
\newcommand {\eenum} {\end{enumerate}}

\newcommand {\bdesc} {\begin{description}}
\newcommand {\edesc} {\end{description}}

\newcommand {\bdefin}[1]{\begin{definition}
                      \mymarginpar{def:#1}
                      \label{def:#1} }
\newcommand {\edefin}       {\end{definition}}
\newcommand {\rdef}[1]{Definition \ref{def:#1}}

\newcommand {\bax}[1]{\begin{axiom}
                      \mymarginpar{ax:#1}
                      \label{ax:#1} }
\newcommand {\eax}       {\end{axiom}}

\newcommand {\bpro}[1]{\begin{property}
                      \mymarginpar{pro:#1}
                      \label{pro:#1} }
\newcommand {\epro}   {\end{property}}

\newcommand {\bprop}[1]{\begin{proposition}
                      \mymarginpar{prop:#1}
                      \label{prop:#1} }
\newcommand {\eprop}       {\end{proposition}}

\newcommand {\blem}[1]{\begin{lemma}
                      \mymarginpar{lem:#1}
                      \label{lem:#1} }
\newcommand {\elem}   {\end{lemma}}
\newcommand {\rlem}[1]{Lemma \ref{lem:#1}}

\newcommand {\bthe}[1]{\begin{theorem}
                      \mymarginpar{the:#1}
                      \label{the:#1} }
\newcommand {\ethe}   {\end{theorem}}
\newcommand {\rthe}[1]{Theorem \ref{the:#1}}

\newcommand {\brem}[1]{\begin{remark}
                      \mymarginpar{the:#1}
                      \label{rem:#1} }
\newcommand {\erem}   {\end{remark}}

\newcommand {\bproof}{\noindent {\bf Proof.} \ }
\newcommand\squares{\vrule height6pt width7pt depth1pt}
\newcommand {\eproof} {\hfill \squares \\ }
\newcommand {\bcor}[1]{\begin{corollary}
                      \mymarginpar{cor:#1}
                      \label{cor:#1} }
\newcommand {\ecor}   {\end{corollary}}

\newcommand {\basu}[1]{\begin{assumption}
                      \mymarginpar{asu:#1}
                      \label{asu:#1} }
\newcommand {\easu}   {\end{assumption}}

\newcommand {\rfig}[1]{Fig. \ref{fig:#1}}

\newcommand {\beq}[1]{\mymarginpar{eq:#1}
                      \begin{equation}
                      \label{eq:#1} }

\newcommand {\beqno}[1]{\mymarginpar{eq:#1}
                      \begin{eqnarray}
                      \nonumber}

\newcommand {\eeq}       {\end{equation}}
\newcommand {\eeqno}       { && \end{eqnarray}}
\newcommand {\req}[1]{(\ref{eq:#1})}

\newcommand {\bear}[1]{\mymarginpar{eq:#1}
                       \begin{eqnarray}
                       \label{eq:#1} }

\newcommand {\bearno}[1]{\mymarginpar{eq:#1}
                       \begin{eqnarray}
                       \nonumber}

\newcommand {\eear}{\end{eqnarray}}
\newcommand {\eearno}{\end{eqnarray}}
\begin{document}

\title{On Sharing Viral Video Over an Ad Hoc Wireless Network}

\author{Yi-Ting Chen, Constantine Caramanis and Sanjay Shakkottai\\
Department of Electrical and Computer Engineering\\
The University of Texas at Austin\\
Email: \{yiting.chen, caramanis, shakkott\}@mail.utexas.edu}

\maketitle

\begin{abstract}
We consider the problem of broadcasting a viral video (a large file) over an ad hoc wireless network (e.g., students in a campus). Many smartphones are GPS enabled, and equipped with peer-to-peer (ad hoc) transmission mode, allowing them to wirelessly exchange files over short distances rather than use the carrier's WAN. The demand for the file however is transmitted through the social network (e.g., a YouTube link posted on Facebook).

To address this coupled-network problem (demand on the social network; bandwidth on the wireless network) where the two networks have different topologies, we propose a file dissemination algorithm. In our scheme, users query their social network to find geographically nearby friends that have the desired file, and utilize the underlying ad hoc network to route the data via multi-hop transmissions. We show that for many popular models for social networks, the file dissemination time scales sublinearly with $n,$ the number of users, compared to the linear scaling required if each user who wants the file must download it from the carrier's WAN.
\end{abstract}

\bsec{Introduction}{intro}

The proliferation of mobile devices that can stream video (laptops, smartphones, tablets) has marked a dramatic increase in demand for streaming video. At the same time, content generation and dissemination has become dramatically easier -- most phones have installed video-cameras, and knowledge of a video can spread extremely rapidly to vast numbers of people, through social networks including e-mail, Facebook, Twitter, and the like. As deployed capacity approaches saturation, we need new transmission architectures to guarantee our wireless networks continue to deliver traffic effectively and efficiently.

This paper addresses precisely this problem. More specifically: we consider the simple, yet increasingly common setting, where a user (e.g., a student on a college campus) generates a large file (a short video, for example) and wants to spread it to her social network -- her friends, their friends, and so on. In the current paradigm, the file creator uploads the file to a central server (e.g., YouTube) and then spreads word of its existence through Facebook, Twitter, etc. Upon learning of the file's existence, interested (we call them ``eager'') users then download the file from the server, using their provider's wide area network (WAN). Since the WAN has bounded bandwidth, the file dissemination time will necessarily scale linearly in the number of users who ultimately receive the file. Particularly in a dense setting like a college campus, this inherently limited centralized scheme for file dissemination may be highly suboptimal. The central question in this paper is: how much better can we do?

Increasingly, smartphones and similar technology, are equipped with both GPS and peer-to-peer transmission modes. In dense environments, this opens the possibility of forming a wireless ad hoc network in which users communicate with each other through several hops of short distance transmissions. As shown in Gupta and Kumar's seminal work \cite{GK00}, the spatial capacity of a wireless ad hoc network scales as $\sqrt{n}$ -- a sharp contrast to the fixed capacity of a WAN. While this scaling spatial capacity of ad hoc networks provides a potential way forward, naive implementation presents severe problems that may leave us worse off than the currently implemented WAN solution. We may have severe congestion caused by subsets of users getting a high number of requests, hence resulting in hot-spots in the network. This will occur, for instance, if users request the file from neighbors on their social network, as most social networks exhibit the presence of super-nodes with very high degree. This is particularly true in the broadcast setting we have here, when we expect there to be such hot spots, which can potentially reduce network capacity by a significant factor \cite{SSG08}.

\subsection{Main contributions}
In this paper we propose a simple and distributed file dissemination algorithm that takes advantage of two main ideas: (i) knowledge of the file spreads quickly because of the structure of the social network -- we can use the same to manage file dissemination; (ii) in dense settings where ad hoc networks make sense, exploiting geographic proximity can provide additional benefits. With these ideas in mind, we devise a file dissemination algorithm that works by passing messages through the social network, and requires limited communication and computation overhead. In particular, the main features of our algorithm are as follows:
\benum{}
\item Load balancing: users receiving a large amount of requests distribute them to nearby users on the social network, in such a way that we can guarantee no user has to serve more than six other users. Our algorithm achieves $\sqrt{n}$-scaling with the number of users receiving the file -- sublinear, in sharp contrast to the linear scaling required in the WAN file dissemination architecture. 

\item Exploiting geographic proximity: We extend our load-balancing algorithm to exploit geographic proximity. Because of the structure of the social network, we show that by searching a few hops deeper in their social network, most users are able to download the file from another user at close range. This idea allows us to further reduce the scaling below $\sqrt{n}$, depending on the depth of the social-network a user may search.

\item Social Networks: We analyze our algorithm on popular models for social networks (power law graphs). We show that the file dissemination time scales sublinearly with $n$ for a broad range of social-network parameters. In addition, we show that the performance of our algorithm is comparable to the best possible dissemination time of any algorithm -- even those not constrained by communication or computation time.
\eenum

\subsection{Related work}
Single piece file dissemination problems were first studied in \cite{FG85}\cite{Pit87}. In \cite{KSS00}-\cite{DM04}, they provide analytic results for multi-piece file dissemination problems. Other topics related to influence spreading, epidemics, and content distribution in social networks can be found, for example, in \cite{KJT03}-\cite{ICM09} and references therein.

Multi-hop transmission in a wireless network has been studied extensively since Gupta and Kumar's seminal work \cite{GK00}. Subsequently, \cite{KV04} provides a simple proof and \cite{FDTT07} closes the gap of $1/\sqrt{\log(n)}$. Multicast and broadcast capacities are considered in, e.g., \cite{LTO07}-\cite{SLS07}. On the other hand, \cite{TE92}-\cite{SSG08} use randomized schemes to balance the traffic load and achieve throughput optimal routing.

\subsection{Paper organization}
We introduce the system model in \rsec{sys}. In \rsec{alg}, we present our algorithm and main results. Some lemmas regarding random placement and random graphs are included in \rsec{lem}. We analyze the performance of our algorithm in \rsec{analysis}. Conclusions are provided in \rsec{con}. The proofs of various lemmas and theorems in \rsec{alg} and \rsec{lem} can be found in Appendix A and Appendix B.

\bsec{System description}{sys}
In this section we describe the basic system model, including the model for the wireless network and the placement of the nodes, and the model for the social network.

\subsection{Random wireless network and Gaussian channel model}
We model our network as $n$ static nodes, placed independently and uniformly on a square of width $\sqrt{n}$. Thus the (expected) density of the network stays constant. Each node has a transmitter and a receiver. All nodes can communicate with each other with fixed power $P$. The interference model is described by a Gaussian channel model defined below. 

\bdefin{def3}(Gaussian channel model) Index nodes by 1, 2, $\ldots, n$. Let $x_i$ be the location of node $i$. Let $\mathcal{A}$ be the set of active transmitters at this time instant. The transmission rate $R(x_i, x_j)$ from node $i$ to node $j$ is
\beq{a1}
R(x_i, x_j) = \log\left(1+\frac{P\ell(x_i, x_j)}{N_0+\sum_{k\in \mathcal{A}\setminus
\{i\}}P\ell(x_k, x_j)}\right).
\eeq
Here, $\ell(x,y)$ represents the power attenuation function between points $x$ and $y$ on the square, and is given by
\beq{a2}
\ell(x_i, x_j) = \min\left\{1, \frac{e^{-\gamma ||x_i-x_j||}}{||x_i-x_j||^\alpha}\right\}
\eeq
where as usual, $||x-y||$ is the Euclidean distance between $x$ and $y$.
\edefin
In this paper, we consider either $\gamma>0$ or $\gamma= 0$ and $\alpha >2$.

\subsection{Model for social networks}
As we identify users with their devices (e.g. cell phones/ PDA), the $n$ nodes in the wireless network also form a social network. A social network is described as a graph $G = (V, E)$ where $V$ is the set of nodes with cardinality $n$ and $E$ is the set of edges. Two nodes are joined by an edge if (and only if) the corresponding users are friends in the social network. The distance between two nodes $x$ and $y$ on the social-graph $G$ is the minimum number of hops between  $x$ and $y$ in the social network. Thus a node's neighbors are the nodes one hop away on the social graph, and its $k$-neighborhood are the nodes within $k$ hops away on the social graph. A key property we exploit is that distance between two nodes on the social network is generally unrelated to geographic distance between the corresponding users in the wireless network.

Empirical studies of many social (and other) networks have shown them to satisfy so-called power law graph structure, including many collaboration networks, but also the Internet and many communication networks (see e.g. \cite{AJB99} \cite{FFF99} \cite{BJN02} \cite{GIC03}). As a consequence, power law graphs (which we define below) are a popular choice for modeling social networks. 

A graph $G$ is called a power law graph with parameter $\beta$ if the number of nodes with degree $k$ is proportional to $k^{-\beta}$. We will consider social networks generated by random power law graphs \cite{CL02}. These random graphs satisfy an important property: each node has only small number of neighbors, i.e., small degree (small relative to the size of the overall network) while the diameter of the random graph (the maximum number of hops between the vast majority of the nodes) is still small, with overwhelming probability. This property is consistent with properties of most social networks, and in particular, with the famous observation known as the small world phenomenon, first discussed in \cite{M67}. 

As is common, we generate random graphs and in particular random power law graphs, according to expected degree sequences \cite{CL02}.
\bdefin{def1} (\cite{CL02})
Let $w = (w_1, w_2, \ldots, w_n)$ be an expected degree sequence satisfying $\max\{w_k^2\}\le \sum_{1\le k \le n}w_k$. We say $G = (V, E)$ is a random graph generated by the degree sequence $w$ if edge $(i, j)\in E$ is present with probability $w_iw_j/\sum_{1\le k \le n}w_k$.
\edefin

\bdefin{def2} (\cite{CL02})
A random graph generated by \rdef{def1} is a random power law graph with parameter $\beta$, average degree $\bar{d}$ and maximum expected degree $M$ if $w_i$ is chosen by
\beq{weight1}
w_i = c(i_o+i)^{-1/(\beta-1)},
\eeq
where $c = \frac{\beta-2}{\beta-1}\bar{d}n^{1/(\beta-1)}$ and $i_0 = n\left(\frac{\bar{d}(\beta-2)}{M(\beta-1)}\right)^{\beta-1}$. 

The well-known Erd\"os-R\'enyi graph, denoted by $G(n, p)$, is the graph where each edge is present with probability $p$. It is thus a random graph with expected degree sequence $w = (np, np, \ldots, np)$.
\edefin

For convenience, we further introduce the following notation. Given a subset $S \subseteq V$, let the volume of $S$ be ${\rm vol}(S) = \sum_{i\in S}w_i$, $i.e.$, the sum of weights of nodes in $S$. Similarly, define ${\rm vol}_k(S) = \sum_{i\in S}w_i^k$ and $\tilde{d} = {\rm vol}_2(G)/{\rm vol}(G)$.
 
\subsection{Assumption on file length}

The transmission time consists of two parts: propagation delay and file receiving time. The propagation delay is the time required to receive the first bit since the start of the transmission. The file receiving time is the time required to finish the transmission since then. For simplicity, we assume the file length $F$ is large, and we ignore the propagation delay in the analysis. 
We note in passing that we can formally incorporate both propagation delay as well as the file receiving time in our analysis by scaling $F$ such that the propagation delay terms will be sub-dominant to the file receiving time.

\bsec{Algorithm and main results}{alg}
We are now ready to present our algorithm and state our main results. At some initial time, the file generator ({\em the source}) creates the file, and advertises it on her social network. At any given time, a node either has the file ({\em active node}), knows about the file and wants it because one of its social-network neighbors has it ({\em eager node}), or is oblivious to its existence ({\em inactive node}). 

The algorithm proceeds in three phases. In the Requesting Phase, eager nodes use their social network to request the file from active nodes -- if knowledge of geographic location is available, nodes favor (geographically) nearby active nodes. In the Scheduling Phase, again the social network is used to schedule a sequence of transmissions whereby each eager node is assigned a transmission node from which it will obtain the file. In the Transmission Phase, nodes transmit the file to their appointed requestors, employing established routing techniques \cite{FDTT07}. This final third phase is conceptually distinct from the first two phases, and it is important to emphasize this point here. The routing techniques used are independent of the social network structure, and follow the multi-hop ad hoc network protocols described in, e.g., \cite{GK00,FDTT07}. Thus, while the requesting and scheduling in Phases 1 and 2 are constrained by the social network, the routing in Phase 3 is not.

We present a single algorithm that accommodates two settings: in the first, simpler setting, nodes have no notion of geography, and may not request the file from active nodes more than a single hop away on their social network. In the second setting, nodes are aware of geography and hence distance, and ``prefer'' to request the file from geographically nearby nodes. Moreover, they are allowed to search for such nearby nodes beyond their immediate neighbors in the social network.

Our algorithm accommodates both settings -- the first, by adjusting the ``preferred distance'' to infinite and the number of search-hops to 1, and the second, by limiting the preferred distance, and by expanding the number of allowed search-hops. In Section \ref{ssec:loadbalancing} we consider the first setting: no geographic information available. We show that for most social networks, our algorithm gives $\sqrt{n}$-scaling. We consider the second setting in Section \ref{ssec:geography}, where nodes have access to geographic position information. We show that again for many social networks, the dissemination time can be further reduced to scale more slowly than $\sqrt{n}$. 

\subsection{Algorithm}
\label{ssec:algorithm}
Our algorithm takes the input as the diameter of the social network, $D$, as well as two parameters which we specify: $\epsilon$, and $\mathcal{L}$, whose roles are as follows. Nodes are allowed to search for another node in the social network from which to download the file, at a distance of at most $2 \epsilon D + 1$ hops away. Thus if $\epsilon = 0$, they cannot look beyond a single hop away, and if $\epsilon = 0.5$, they have access to the entire social network. Thus the parameter $\epsilon$ controls the search depth. The parameter $\mathcal{L}$ is used to exploit geographic proximity: most nodes will download the file from nodes that are at a geographic distance of at most $\mathcal{L}$. If nodes have no notion of geography, we set $\mathcal{L}=\infty$, hence all nodes are within $\mathcal{L}$. Otherwise, we set $\mathcal{L}$ to a smaller value.


Given parameters $(\epsilon, \mathcal{L}, D)$ as described above, the algorithm finds active nodes from which eager nodes can download the file. This is accomplished through coordination through the social network. 

The main idea is the following: eager nodes send requests to one of their social-network neighbors with the file. Since a single node may get many such requests, it does not serve all of them, but rather finds other active nodes nearby in the social network to serve them, and also enlists the receiving nodes themselves to forward along the file. The theorems given in Sections \ref{ssec:loadbalancing} and \ref{ssec:geography} show that for the specific choices of parameters $\epsilon$ and $\mathcal{L}$ given, the algorithm succeeds in delivering the file to all nodes, and moreover does so in the advertised time scaling.

When $\mathcal{L}$ is set to a non-infinite value, it may not always be possible for nodes to obtain the file from geographically proximate neighbors -- for instance, suppose the generator has no neighbors in her geographic proximity. In such cases, we allow file transfers that exceed geographic distance $\mathcal{L}$, and these happen from two or one-hop neighbors on the social network. We call transfers within geographic distance $\mathcal{L}$, $\mathcal{L}$-transfers, and all other transfers $S$-transfers, since they are near in the social-network distance. Similarly we refer to ${\mathcal{L}}$-requests and $S$-requests.
%
%

\textbf{ALGORITHM 1:}

Input: parameter $\epsilon$, distance threshold $\mathcal{L}$, and the diameter of the social network $D$.

\textit{Requesting Phase:} Consider an eager node, $x$, at time $t$. 

Step 1: Let $\mathcal{N}_x(t)$ denote node $x$'s $2\epsilon D+1$-neighborhood in the social-graph at time $t$. Let $\mathcal{N}_x^{\mathcal{L}}(t) \subseteq \mathcal{N}_x(t)$ be the set of nodes in $\mathcal{N}_x(t)$ that have the file and whose Euclidean (geographic) distance to $x$ does not exceed $\mathcal{L}$.

Step 2: If $\mathcal{N}_x^{\mathcal{L}}(t)$ is not empty, $x$ sends an ${\mathcal{L}}$-request to a randomly picked node in $\mathcal{N}_x^{\mathcal{L}}(t)$. 

Step 3: If $\mathcal{N}_x^{\mathcal{L}}(t)$ is empty and the distance from $x$ to the source on the social-graph is smaller than $\epsilon D+1$, then $x$ sends an $S$-request to a one-hop neighbor in the social-graph which has the file.

Step 4: Otherwise, $x$ waits and goes back to step 1 at time $t+1$.

\begin{figure}[htb]
        \center{\includegraphics[width = 35 mm]
        {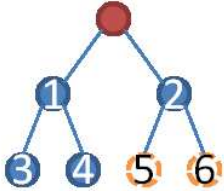}}
        \caption{\label{fig:binary} Each active node maintains balanced binary trees and adds requesting nodes to trees sequentially. Suppose the active node depicted at the root gets four requests at time $t$, and two more at time $t+1$. The resulting tree might look as depicted. The original active node would then serve nodes 1 and 2, subsequently node 1 would serve nodes 3 and 4, and node 2 would serve nodes 5 and 6.}
\end{figure}

\textit{Scheduling Phase:} Consider an active node $y$. 
It maintains two balanced binary trees, an $\mathcal{L}$-tree and an $S$-tree, constructed from its $\mathcal{L}$-requests and $S$-requests, respectively. It builds these trees by adding requesting nodes sequentially, as the requests arrive. This sequential building of the binary trees is depicted in Figure \ref{fig:binary}.

When node $y$ receives an $\mathcal{L}$-request, node $y$ adds the eager node to the $\mathcal{L}$-tree and asks its parent on the tree to deliver the file, and similarly for $S$-requests.
%
%
%

\textit{Transmission Phase:} An eager node waits until the node designated as its transmitting node in the Scheduling Phase has the file. It then sets up a wireless transmission, and routes data through a highway system described in \cite{FDTT07}. Note that the transmitter will have to serve at most 6 nodes: 2 from its own $\mathcal{L}$-tree, 2 from its own $S$-tree, and 2 from the tree it joins when it is an eager node (which could be either an $\mathcal{L}$-tree or an $S$-tree). Thus, we divide a time slot into six and each transmitter serves all nodes in a round robin fashion. 
\subsection{Main results: load balancing}
\label{ssec:loadbalancing}
In this section we show that the load-balancing accomplished by the $\mathcal{L}$-binary trees is enough to give $\sqrt{n}$-scaling, without any geographic information. We show that our result holds, as long as the social network has the properties of a random power law graph with $\beta >2$, minimum expected degree $m>3$ and maximum expected degree $M$ satisfying $\log(n) \ll M\ll \sqrt{n}$ (many social networks have values of $\beta$ large than this -- see, for example, collaboration graphs in \cite{BJN02}). In this case, the diameter of the social-graph is $\mathcal{O}(\log(n))$ and the size of the largest component is of $\Theta(n)$ \cite{CL02}\cite{CL02B}.

As discussed, we set $\mathcal{L}=\infty$, and $\epsilon = 0$, thus nodes are only allowed to request the file from nodes at most one hop away on the social network, and they entirely ignore geography. 

In this case, for any eager node $x$, we have $\mathcal{N}_x^\mathcal{L}(t) \neq \emptyset$ at the time $t$ node $x$ becomes eager, and hence the Requesting Phase of the algorithm uses only Step 1 and Step 2. There are only $\mathcal{L}$-requests, and thus the algorithm requires each node to transmit to at most 4 other nodes. Indeed, the point of this algorithm is to distribute the load evenly on the wireless network.

In \rthe{flooding}, we show that the file dissemination time scales like $\sqrt{n}$ (sublinearly). In addition, we show that the performance only differs from {\it algorithm independent} lower bounds with a factor $n^\xi$ for any $\xi>0$. Since the proofs of the following two theorems are similar to those for \rthe{neighborsearch} and \rthe{PLG-lower2}, we defer the full details to the Appendix B.

\bthe{flooding} 
Consider the file dissemination problem with wireless network and social network as defined above. Suppose the file length is $F$. Then the file dissemination time for Algorithm 1 is 
\beq{ccc-1}
\mathcal{O}(\sqrt{n}\log^2(n)F)
\eeq
with high probability.
\ethe

\bthe{PLG-lower1}
Consider the file dissemination problem with wireless network and social network as defined above. Suppose the file length is $F$. Then, for any algorithm that allows nodes to download the file from their 1- and 2-hop neighbors on the social network, the file dissemination time is lower bounded by
\beq{ccc0}
\Omega(n^{1/2-\xi}F),
\eeq
for any $\xi>0$ with high probability.
\ethe

\brem{r1}
Significantly, the only properties of power law graphs we use are the size of the diameter and the maximum degree. Specifically, given a graph $G$ with diameter $\ell_{max}$ and maximum degree $d_{max}$, the file dissemination time is $\mathcal{O}(\sqrt{n}\log(d_{max})\ell_{max}F)$ if nodes are only allowed to download the file from nodes at most 2 hops away. The proof of this follows immediately from the proof of the theorem.
\erem

\subsection{Main results: exploiting geography}
\label{ssec:geography}
Intuitively, increasing the number of geographically proximal downloads should decrease transmission time. We show that this can be accomplished, at the cost of deeper searching of the social network, as long as the social network has the properties of a random power law graph with $\beta >3$ (again, many graphs have this property, see, e.g., the collaboration graphs in \cite{GIC03}). We assume that the minimum expected degree is $m = K\log(n)$ where $K$ is a constant greater than 10, and the maximum expected degree is $M$, satisfying $\log^2(n) \ll M\ll \sqrt{n}.$ Thus, almost all nodes are in the largest component and the diameter of the graph is $D \approx \log_{\tilde{d}}(n)$ \cite{CL02}\cite{CL02B} (recall the definition of $\tilde{d}$ from Section II). 


Setting $\epsilon$ to a positive value translates to allowing nodes to search for an active node in their $2\epsilon\log_{\tilde{d}}(n)+1$-neighborhood, and because of our load-balancing architecture, ultimately download the file from nodes in their $4\epsilon\log_{\tilde{d}}(n)+2$ neighborhood. With more active nodes available, eager nodes can more easily find geographically proximal active nodes. We set $\mathcal{L} = 8\sqrt{n^{1-\epsilon'}\log(n)/\sigma\pi}$ for any $\epsilon'<\epsilon$. The value of $\epsilon$ is chosen to be small, $\epsilon < 1/10$, allowing nodes to search a neighborhood that is large, but nevertheless a vanishing fraction of the size of the entire network. 
%

As load-balancing alone was able to achieve file dissemination time scaling of $\sqrt{n}$, we show now that by additionally exploiting geography, the file dissemination time can be further reduced by a factor $n^{\epsilon/2}$ compared to the result in \rthe{flooding}. Proofs of the two theorems can be found in \rsec{analysis}.

\bthe{neighborsearch}
Suppose the source is chosen uniformly at random from the nodes in the largest component and the file length is $F$. Consider the setting described above. Then the file dissemination time under Algorithm 1 with parameter $0<\epsilon<0.1$ is 
\beq{g1}
\mathcal{O}(\sqrt{n^{1-\epsilon'}}\log^{2.5}(n)F),\nonumber
\eeq
for any $\epsilon'<\epsilon$ with high probability.
\ethe

\bthe{PLG-lower2}
Consider the file dissemination problem under the setting described above. Let $F$ be the file length. Then, for any algorithm that allows nodes to download the file from their $4\epsilon\log_{\tilde{d}}(n)+2$-neighborhood with $\epsilon<0.1$, the file dissemination time is lower bounded by
\beq{zzz0}
\Omega(n^{1/2-2\epsilon-\xi}F),
\eeq
with high probability for any $\xi>0$.
\ethe

\bsec{Random Placement and Random Graphs}{lem}
In preparation for the proof in the next section, we give some lemmas that characterize the behavior of randomly placed nodes in a square, and also give properties of random graphs.

\subsection{Results about random placement of nodes in a square}
Two properties in particular, are important. For our scheme to work, we need to show that with overwhelming probability, we will not have a very high clustering of nodes (some clustering will occur). We also need to show that when nodes look in their social network for geographically proximate active nodes, they will be able to find at least one, with high probability. The next two lemmas show precisely these properties. 

In the first lemma, we show we control the minimum distance between a node and $k$ other nodes. We use this lemma to ensure that each node can find a node close to it on the wireless-square. In the second lemma, we show a concentration result about the number of nodes falling into a small rectangle, thus showing it is not too big. The proofs of these lemmas are also available in Appendix B.

\blem{dist}
Place $k+1$ nodes on a square of width $\sqrt{n}$ independently and uniformly. Let $\tau$ be the minimum distance from the first node to the others. Then, we have
\beq{bbb3}
\mathbb{P}(\tau\ge \sqrt{64n\log(n)/\pi k}) \le n^{-2}.
\eeq
\elem

\blem{rec}
Place $n$ nodes on a square of width $\sqrt{n}$ independent and uniformly. Given a rectangle of area $A$ where $A = \omega(\log(n))$, let $X$ be the number of nodes in the rectangle. Then, 
\beq{bbb4}
\mathbb{P}(X \ge 2A) \le n^{-2}.
\eeq
\elem

\subsection{Results about the neighborhood behavior of random graphs}
In the following lemma, we address the relation between weights and the number of neighbors. Specifically, we show that if a node has weight $w_i$ greater than $10\log(n)$, then the number of one-hop neighbors the node can reach in the social-graph is between $w_i/2$ and $2w_i$. We use this lemma as it provides a relationship between weights and the number of nodes.

\blem{node}
Suppose $w_i \ge 10\log(n)$. Let $X$ be the number of one-hop neighbors in the social-graph of node $i$. Then, $w_i/2 < X < 2w_i$ with probability $1-o(n^{-1}).$
\elem

The next two lemmas characterize the local behavior of random power law graphs. Specifically, we are interested in how the size of neighborhoods of nodes in the largest component grows. We show that for any node in the largest component, the number of nodes in a small neighborhood grows like a factor $\tilde{d}$ if we explore one more step. We prove this by providing upper and lower bounds that only differ by a factor of $n^\xi$ for any $\xi>0$. The proofs are shown in Appendix A.

\blem{PLG-lower}
Consider a random power law graph with parameter $\beta>3$. Suppose the minimum expected degree is $m = K\log(n)$ for some $K \ge 10$ and the maximum expected degree is $M \gg \log^2(n)$. Then, there are at least $\sigma n^{\epsilon'}$ nodes in a node's $\epsilon\log_{\tilde{d}}(n)$-neighborhood with probability $1-o(n^{-1})$, for any $\epsilon'<\epsilon<0.1$. Here, $\sigma$ is a constant depending on $\beta$ and $K$.
\elem

\blem{PLG-upper}
Consider a random power law graph with parameter $\beta >3$. Suppose the minimum expected degree is $m = K\log(n)$ for some $K\ge 10$, the maximum expected degree is $M \gg \log^2(n)$, and $\epsilon <0.4$. Consider a node either picked randomly or with weight smaller than $W$. Then, there are at most $2W\tilde{d}^\lambda n^{\epsilon'}/\log(n)$ nodes in this node's $\epsilon\log_{\tilde{d}}(n)+\lambda$-neighborhood, with probability $1-\mathcal{O}(\log^{-1}(n))$,  for any $\epsilon'>\epsilon$ and any fixed constant $\lambda$ where 
\beq{ssss1}
W =\left\{ \begin{array}{ccc}
\log^{\beta/\beta-3}(n) & \mbox{if} & 3 < \beta \le 4  \\ 
\max\{\log^{5/\beta-4}(n), \log^2(n)\} & \mbox{if} & 4 < \beta
\end{array}\right.
\eeq
\elem

\bsec{Performance Analysis}{analysis}
\subsection{Proof of \rthe{neighborsearch}}
In this section, we first prove \rthe{neighborsearch} which states the performance of our algorithm when geographic information is available, and when nodes can download the file from a  neighborhood of radius $4\epsilon\log_{\tilde{d}}(n)+2$. The proof of the more simple load-balancing case (where we set $\mathcal{L}=\infty$ and $\epsilon = 0$) is essentially a consequence of this proof -- for the full details we refer to Appendix B. Specifically, we show the file dissemination time is roughly $\sqrt{n^{1-\epsilon}}F$.

\begin{figure}[htb]
        \center{\includegraphics[width = 35 mm]
        {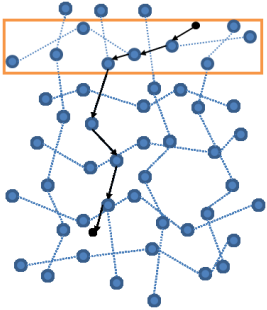}}
        \caption{\label{fig:routing2} An illustration of the highway system and routing. Packets are first routed through horizontal highways to vertical highways corresponding to destinations.}
\end{figure}

The proof of the theorem consists of two main parts: showing the existence of a geographically nearby neighbor in the wireless-graph and the analysis of transmission rates. In addition, the transmission phase of our algorithm relies on some routing results from \cite{FDTT07} and \cite{LLL08} which we summarize here. For the full details, we refer readers to those individual papers. In the routing scheme, packets are routed through a highway system consisting of horizontal highways and vertical highways. Each highway serves nodes in a stripe on the wireless-square. An illustration is shown in \rfig{routing2}. The results in \cite{FDTT07} and \cite{LLL08} guarantee the following properties of this highway system.
\benum{}
\item Nodes can reach their highways in a hop of length $\mathcal{O}(\log(n))$.
\item The highways are almost straight. For example, if a flow on a horizontal highway starts from $x$-coordinate $a_1$ with destination at $x$-coordination $a_2 > a_1$, it will not reach any node with $x$-coordinate smaller than $a_1-H$ where $H = \mathcal{O}(\log(n))$.
\item Highway nodes can communicate with neighboring highway nodes with a constant rate. A highway node serves flows through it with equal rate.
\eenum
We now move to the proof of the theorem. We first state the existence of an ``intermediate node'' in the following lemma.  

\blem{existence}
Consider a random power law graph with $\beta>3$ and $m = K\log(n)$ where $K>10$ is a constant. For some $\epsilon<0.1$, consider a node $x$ in the largest component, such that the distance from $x$ to the source on the social network is greater than $\epsilon\log_{\tilde{d}}(n)$. Then, there exists a node $y$ which satisfies the follows with probability at least $1-o(n^{-1})$:
\benum{}
\item $y$ is in the $2\epsilon\log_{\tilde{d}}(n)+1$-neighborhood of $x$ in the social-graph.
\item The distance from $y$ to the source on the social network is smaller than that from $x$ to the source.
\item The Euclidean distance from $x$ to $y$ on the wireless-square is smaller than $\mathcal{L}$.
\eenum
\elem
\bproof
We first show that there exist $\sigma n^{\epsilon'}$ nodes satisfying 1) and 2) with probability $1-o(n^{-1})$. Let $d_x$ be the distance from $x$ to the source on the social network. Since $d_x > \epsilon\log_{\tilde{d}}(n)$, there exists a node $z$ such that the distance from $z$ to $x$ on the graph is $\epsilon\log_{\tilde{d}}(n)+1$ and the distance from $z$ to the source on the graph is $d_x-\epsilon\log_{\tilde{d}}(n)-1$. Therefore, nodes in the $\epsilon\log_{\tilde{d}}(n)$-neighborhood of $z$ in the social-graph satisfy 1) and 2). In addition, by \rlem{PLG-lower}, the size of such a neighborhood is greater than $\sigma n^{\epsilon'}$ with probability $1-o(n^{-1})$.

Thus, by \rlem{dist}, there exists a node $y$ among the $\sigma n^{\epsilon'}$ nodes whose Euclidean distance to $x$ on the wireless-square is smaller than $\mathcal{L}$ with probability $1-o(n^{-1})$.
\eproof

\bproof (\rthe{neighborsearch})
Recall that our algorithm classifies transmissions as those chosen because they are geographically within distance $\mathcal{L}$, called $\mathcal{L}$-transmissions, and those chosen because they are within two hops on the social network, called $S$-transmissions. $S$-transmissions are those whose Euclidean distances between transmitters and receivers on the wireless-square are not guaranteed to be less than $2\mathcal{L}$, as are $\mathcal{L}$-transmissions. Note that the number of $S$-transmissions is smaller than the number of nodes in $\epsilon\log_{\tilde{d}}(n)$-neighborhood of the source in the social-graph which is smaller than $2Wn^{\epsilon''}/\log(n)$ for any $\epsilon''>\epsilon$ with high probability by \rlem{PLG-upper}.

Now, we bound the number of flows through a highway node at any time. Consider a transmission between two nodes with Euclidean distance less than $2\mathcal{L}.$ By the fact that highways are almost straight and the first and last hops are of length $\mathcal{O}(\log(n))$, the transmission passes through a horizontal (vertical) highway node only if the horizontal (vertical) distance between the transmitter (receiver) and the node is smaller than $3\mathcal{L}$ on the wireless-square. In other words, $\mathcal{L}$-transmissions through a horizontal (vertical) highway node must fall in a rectangle of side $6\mathcal{L}\times h$ in the corresponding horizontal (vertical) strip where $h$ is a constant provided in \cite{FDTT07}. Since, by \rlem{rec}, the total number of nodes falling into this region is $\mathcal{O}(\mathcal{L})$ with probability $1-o(n^{-1})$ and each node generates at most a constant number of flows, using the union bound we can conclude that all highway nodes have at most $\mathcal{O}(\mathcal{L})$ $\mathcal{L}$-flows with probability $1-o(1)$. In addition, since there are at most $2Wn^{\epsilon''}/\log(n)$ $S$-transmissions, the total number of flows through each highway node is $\mathcal{O}(\mathcal{L})$ with probability $1-o(1)$. Therefore, each flow has a rate $\Omega(1/\mathcal{L})$ with high probability and each node can receive the file in $c_0\mathcal{L}F$ time slots for some constant $c_0>0$ from the time when the transmission begins.

We prove the theorem by induction on $k$: the distance from a node to the source on the social-graph. Let $\mathcal{N}_k$ denote nodes whose distance to the source is $k$ on the social-graph. The claim of the induction is that a node in $\mathcal{N}_k$ can receive the file in at most $kc_0\log_2(n)\mathcal{L}F$ time slots. By our notation, $\mathcal{N}_0$ is the source node. First note, that the base case $k=1$ of the induction clearly holds. 
Now, we suppose it is true for $k-1$ and consider nodes in $\mathcal{N}_k$. Note that no nodes in $\mathcal{N}_k$ are inactive at time $(k-1)c_0\log_2(n)\mathcal{L}F$. Further, by Algorithm 1 and \rlem{existence}, all nodes in $\mathcal{N}_k$ can request the file, according to the algorithm, from an active node in $\cup_{i=0}^{k-1}\mathcal{N}_i$. Thus, these nodes have to wait at most $\log_2(n)-1$ successful transmissions before starting to receive the file, since the depth of any binary tree is at most $\log_2(n)$. Therefore, they can receive the file before time $kc_0\log_2(n)\mathcal{L}F$. Hence, by induction, the file dissemination time is $\mathcal{O}(\sqrt{n^{1-\epsilon'}}\log^{2.5}(n)F)$ as the diameter of the social-graph is $\mathcal{O}(\log(n))$.
\eproof

\subsection{Proof of \rthe{PLG-lower2}}
We proceed by first providing some definitions and a lemma. Given a transmission pair with rate $r$ over an Euclidean distance $\rho$ on the wireless-square, define the bit-meter rate of the transmission pair as $r\rho$. The total bit-meter product a network can transmit is the supremum of the sum of bit-meter products of all transmission pairs.

\blem{bmlimit}
The total bit-meter product the network can transmit in a time slot is $\Theta(n)$.
\elem


\bproof
From \req{a1}, we know the bit-meter product a transmission pair $(x_i, x_j)$ can transmit is
\bear{sss1}
&&||x_i-x_j||\log\left(1+\frac{P\ell(x_i, x_j)}{N_0+\sum_{k\neq i}P\ell(x_k, x_j)}\right)\nonumber\\ 
&&\le P\ell(x_i, x_j)||x_i-x_j||/N_0.
\eear
Recall that $\ell(x_i, x_j)||x_i-x_j||$ is bounded by a constant either for $\gamma >0$ or $\gamma = 0$ and $\alpha > 2$. Since there are at most $n/2$ transmission pairs, the total bit-meter product the system can transmit is $\Theta(n)$ in a time slot.
\eproof

To prove the lower bound, we place no restrictions on computation or communication overhead. Moreover, we make (overly) optimistic assumptions throughout in order to guarantee a bound. For instance, we assume nodes only download from their nearest social-network neighbors.


\bproof (\rthe{PLG-lower2})
Define the transport load as the infimum of the total bit-meter product required to disseminate the file under the problem setting. To apply \rlem{bmlimit}, we just need to show that the transport load is $\Omega(n^{3/2-2\epsilon-\xi}F)$ with probability $1-o(1)$.

Let $\mathcal{M}$ be the set of nodes in the largest component with expected degree in the range $[K\log(n)$ $2K\log(n)]$. Then, 
\beq{ccc1}
|\mathcal{M}| \approx \frac{\int_{K\log(n)}^{2K\log(n)}x^{-\beta}ndx}{\int_{K\log(n)}^Mx^{-\beta}dx} = (1+o(1))(1-2^{1-\beta})n.
\eeq
since almost all nodes are in the largest component.

Fix any $\epsilon' >\epsilon$. Let $\mathcal{N}_i$ be the set of nodes that node $i$ can reach in $4\epsilon\log_{\tilde{d}}(n)+2$ hops in the social-graph. Let $X_i$ be the indicator that the Euclidean distance from node $i$ to $\mathcal{N}_i$ on the wireless-square is smaller than $\sqrt{n/2\pi W\tilde{d}^2n^{4\epsilon'}}$. Therefore, we have, for $i \in \mathcal{M}$ and $n$ large enough and some constant $c_1$,
\bear{ccc2}
\mathbb{P}(X_i = 1) &\le& \mathbb{P}(\{X_i = 1\} \cap \{|\mathcal{N}_i|\le 2W\tilde{d}^2n^{4\epsilon'}/\log(n)\}) \nonumber\\ 
&&+ \mathbb{P}(|\mathcal{N}_i|> 2W\tilde{d}^2n^{4\epsilon'}/\log(n)) \nonumber\\
&\le& 1/\log(n) + \mathcal{O}(1/\log(n)) \le c_1/\log(n)\nonumber
\eear
since the probability that a node is close is smaller than $1/2W\tilde{d}^2n^{4\epsilon'}$ and the second term comes from the probability that $|\mathcal{N}_i|\ge 2W\tilde{d}^2n^{4\epsilon'}/\log(n)$. Therefore, we have
\beq{ccc3}
\mathbb{E}\left[\sum_{i\in \mathcal{M}}X_i\right] \le 2c_1(1-2^{1-\beta})n/\log(n)
\eeq

We claim that $\mathbb{P}(|\sum_{i\in \mathcal{M}}X_i-\mathbb{E}[\sum_{i\in \mathcal{M}}X_i]|\ge n/\log^{1/3}(n)) = o(1).$ Indeed, by Chebyshev's inequality, we have
\bear{ccc7}
&&\mathbb{P}\left(\left|\sum_{i\in \mathcal{M}}X_i-\mathbb{E}\left[\sum_{i\in \mathcal{M}}X_i\right]\right|\ge n/\log^{1/3}(n)\right)\nonumber\\
&&\le \frac{\mathbb{E}[(\sum_{i\in \mathcal{M}}X_i)^2]}{n^2/\log^{2/3}(n)}\nonumber\\
&&\le \frac{c_1n^2/\log(n)}{n^2/\log^{2/3}(n)} = o(1)
\eear
where the last inequality follows from $\mathbb{E}[X_iX_j] \le \mathbb{E}[X_i]$.

By the above claims, $|\mathcal{M}| = \Theta(n)$ while the number of nodes with geographically close neighbors in the wireless-square is $o(n)$. Hence, the transport load is $\Omega(n^{3/2-2\epsilon-\xi}F)$.
\eproof

\bsec{Conclusions}{con}
New technology (smartphones, etc.) has made content creation easy -- just a press of a button. Social networks, meanwhile, make wide dissemination of the {\it knowledge of} that file, just as easy -- a press of another button. Yet actual dissemination of large files to many users can seriously burden a wireless network. In the WAN setting, the time to disseminate must scale linearly in the number of users. In this paper, we consider simple, low-overhead file dissemination algorithm that exploits peer-to-peer capabilities of many smartphones and similar devices, and, critically, exploits the very social networks that spread knowledge of the file. We give a load-balancing algorithm that uses the social network to schedule transmissions so that spatial-capacity of the ad hoc network is exploited without creating congestion or hot spots. We show that dissemination time scales like $\sqrt{n}$ --- significantly slower than the linear time for WAN. Then, we show that if nodes have knowledge of geographic position, this can be exploited to further decrease file dissemination time. Finally, we show in both cases that our algorithm performs close to an algorithm-independent lower bound.
%
%
%
%
%

\bsec{Appendix A}{appa}
\subsection{Proof for \rlem{PLG-lower}}

We first quote lemma 3.2 from \cite{CL02}. This useful lemma addresses how a neighborhood of a set in the random power law graph grows. Specifically, if we have two sets $S$ and $T$, what is the sum of weights of neighbors of $S$ which are also in $T$?  One important application is the setting where $T \approx G$, i.e., $T$ is almost the entire graph. In this case, we get an increase factor of roughly $\tilde{d}$.

\blem{expansion} (\cite{CL02})
Given a random graph and two subsets $S$ and $T$,  if 
\beq{qqq2a}
\frac{2c}{\delta^2}\frac{{\rm vol}_3(T)}{{\rm vol}_2^2(T)}\le \frac{{\rm vol}(S)}{{\rm vol}(G)},
\eeq
\beq{qqq2b}
\frac{{\rm vol}(S)}{{\rm vol}(G)}\le \delta\frac{{\rm vol}_2(T)}{{\rm vol}_3(T)},
\eeq
we have
\beq{qqq1}
{\rm vol}(\Gamma(S)\cap T) \ge (1-2\delta)\frac{{\rm vol}_2(T)}{{\rm vol}(G)}{\rm vol}(S)
\eeq
with probability $1-e^{-c}$ where $\Gamma(S)$ is the set of one-hop neighbors of $S$.
\elem

Using this lemma, we provide a proof to \rlem{PLG-lower}, which we used to lower bound the size of a node's immediate neighborhood. 

\bproof (\rlem{PLG-lower})
Consider node $x$'s neighborhood. Let $S_i$ be the set of nodes whose distance to $x$ is $i$ and $S_0= \{x\}$. We will show that ${\rm vol}(S_{i+1})\ge (1-2\delta)\tilde{d}{\rm vol}(S_i)$ for $\delta = 1/4$ with probability $1-o(n^{-2})$. To do this we need to apply \rlem{expansion} inductively and choose $c = 3\log(n)$. We may assume ${\rm vol}(S_i) < n^{2\epsilon}$ in the first $\epsilon\log_{\tilde{d}}(n)$ steps. Since $\frac{{\rm vol}_2(T)}{{\rm vol}_3(T)}{\rm vol}(G) = \Omega(n/M)$ for all $T$, \req{qqq2b} holds for all $\beta >3$. We have only to verify \req{qqq2a}.

First notice that ${\rm vol}(S_1) = \Omega(\log^2(n))$ with probability $1-o(n^{-1})$. This is true since, by \rlem{node}, the node $x$ has at least $K\log(n)/2$ neighbors with probability $1-o(n^{-1})$ and each neighbor has weight at least $K\log(n)$.

We next verify \req{qqq2a} for $\beta > 4$. Let $T_i$ be the set of all potential nodes whose distance to $x$ is $i+1$, i.e., $T_i = G\setminus\cup_{k = 0}^{i}S_k$. Then, ${\rm vol}_2(T_i) = (1+o(1)){\rm vol}_2(G)$ and ${\rm vol}_3(T_i) = (1+o(1)){\rm vol}_3(G)$. Thus, $\frac{{\rm vol}_3(T_i){\rm vol}(G)}{{\rm vol}^2_2(T_i)} = \Theta(1)$. Therefore, we have the result by induction.

For the case $3< \beta \le 4$, let $T^{(k)}_i$ be the intersection of $T_i$ and the set of nodes with weight smaller than $k\log(n)$. Then, we have
\beq{qqq4}
\frac{2c}{\delta^2}\frac{{\rm vol}_3(T_i^{(k)})}{{\rm vol}^2_2(T_i^{(k)})}{\rm vol}(G) = \mathcal{O}(\log(n))
\eeq
and
\bear{qqq5}
\frac{{\rm vol}_2(T_i^{(k)})}{{\rm vol}_2(G)} \approx \frac{\int_{K\log(n)}^{k\log(n)}x^{2-\beta}dx}{\int_{K\log(n)}^{M}x^{2-\beta}dx} \nonumber\\ 
= (1+o(1))(1-(\frac{k}{K})^{3-\beta}).
\eear
Therefore, by \req{qqq4}, we have \req{qqq2b} is true by induction. On the other hand, ${\rm vol}_2(T_i^{(k)})/{\rm vol}(G) \approx \tilde{d}$ as $k$ becomes large enough, by \req{qqq5}. 

With the above results, we can conclude that the size of the neighborhood grows by roughly a factor of $\tilde{d}/2$ with probability $1-o(n^{-2})$ for each step, from the second one to the $\epsilon\log_{\tilde{d}}(n)^{th}$ step. Since ${\rm vol}(S_{\epsilon\log_{\tilde{d}}(n)-1}) \ge 2\sigma n^{\epsilon(1-o(1))}$ for some constant $\sigma$, there are at least $\sigma n^{\epsilon'}$ nodes within distance $\epsilon\log_{\tilde{d}}(n)$ of $x$ in the social network.
\eproof

\subsection{Proof for \rlem{PLG-upper}}
The proof of \rlem{PLG-upper} depends on the following lemma from \cite{CL02B}, that provides large deviation results for both an upper bound and a lower bound for the sum of Bernoulli random variables.

\blem{bino1} (\cite{CL02B})
Let $X_i$ be a Bernoulli random variable with parameter $p_i$. Suppose $\{X_i\}$ are independent. Let $X = \sum_{i=1}^na_iX_i$ and $\nu = \sum_{i=1}^na_i^2p_i.$ Then, we have
\beq{bbb1}
\mathbb{P}(X\le \mathbb{E}[X]-c)\le \exp(-c^2/2\nu)
\eeq
\beq{bbb2}
\mathbb{P}(X\ge \mathbb{E}[X]+c)\le \exp(-c^2/2(\nu+ac/3))
\eeq
where $a = \max\{a_1, a_2, \ldots, a_n\}$.
\elem

\bproof (\rlem{PLG-upper})
We first state the flow of the proof. In the beginning, we show that we only need to consider an initial node $x$ with weight $W$. We next define $S_i$ as the set of nodes at distance $i$ from node $x$ and show that, for any $\delta>0$,
\beq{asu1}
{\rm vol}(S_i) \le W((1+\delta)\tilde{d})^i.
\eeq
We in fact show ${\rm vol}(S_{i+1}) \le (1+\delta)\tilde{d}{\rm vol}(S_i)$ with probability $1-\mathcal{O}(\log^{-2}(n))$. To do so, we construct a set $T^i$ which contains nodes with large weight and show $T^i\cap S_{i+1} = \emptyset$ with overwhelming probability. On the other hand, we use \rlem{bino1} to bound the sum of weights in $S_{i+1}$ contributed by nodes with small weight. To do this, we have to consider three cases depending on $\beta$. 

We now present the details of the proof. We first show the condition, ${\rm vol}(S_0) \le W$. Since $S_0 = \{x\}$, we need to show the weight of $x$ does not exceed $W$ if $x$ is picked randomly. 
\bear{xxx1}
\int_{\log^2(n)}^{M}x^{-\beta}dx/C_1 &=& \frac{1+o(1)}{C_1(\beta-1)}(\log^2(n))^{1-\beta} \nonumber\\ 
&=& o(\log^{-2}(n)),
\eear
where $C_1$ is the normalization constant $\int_{m}^Mx^{-\beta}dx$. Thus, a randomly picked node $x$ has weight smaller than $W$ with high probability. Since by standard coupling arguments we see that the growth of the neighborhood of $x$ is dominated by a node with weight $W$, we simply take the weight of $x$ to be $W$ in what follows.

We turn to show ${\rm vol}(S_{i+1})\le (1+\delta)\tilde{d}{\rm vol}(S_i)$ with probability $1-\mathcal{O}(\log^{-2}(n))$. First, we give some definitions. Let $\tilde{m}_i = ({\rm vol}(S_i)\log^\beta(n))^{1/(\beta-2)}$ and define $T^i$ to be the set of nodes with weight greater than $\tilde{m}_i$. We first show that $T^i\cap S_{i+1} = \emptyset$ with probability $1-\mathcal{O}(\log^{-2}(n))$. Indeed, 
\bear{xxx3}
\mathbb{P}(S_{i+1}\cap T^i\neq \emptyset) &\le& {\rm vol}(S_i){\rm vol}(T^i)/{\rm vol}(G) \nonumber\\
&\approx& {\rm vol}(S_i)\int_{\tilde{m}_i}^Mx^{1-\beta}ndx/C_1{\rm vol}(G) \nonumber\\
&=& \mathcal{O}(\log^{-2}(n)),
\eear
where the first inequality follows from the union bound. Therefore, with high probability, $S_{i+1}\cap T^i = \emptyset$.

Define $\tilde{T}_i$ to be the set of unexplored nodes with weight smaller than $\tilde{m}$ in the $i$-th step, i.e., $\tilde{T}_i = V\setminus (T^i \cup \cup_{j = 0}^i S_j)$. Thus, $S_{i+1}$ is a subset of $\tilde{T}_i$. For simplicity, we consider a larger set $T_i$ which includes $\tilde{T}_i$ and virtual nodes $\tilde{V}_i$ where $\tilde{V}_i$ is chosen such that the number of nodes and their weights are the same as those in $\cup_{j = 0}^i S_j$. We allow nodes in $S_i$ to connect to nodes in $\tilde{V}_i$ and, therefore, have a looser upperbound on ${\rm vol}(S_{i+1})$. We first give some properties of $T_i$. Specifically, we claim ${\rm vol}(T_i) = (1+o(1)){\rm vol}(G)$ and ${\rm vol}_2(T_i) = (1+o(1)){\rm vol}_2(G)$. Indeed,  
\bear{xxx4}
{\rm vol}(T_i) &=& {\rm vol}(G) - {\rm vol}(T^i) \nonumber\\
&\ge& {\rm vol}(G) - \int_{\log^{\beta/(\beta-2)}(n)}^Mx^{1-\beta}ndx/C_1\nonumber\\
&=& {\rm vol}(G) - \mathcal{O}(n\log^{-1}(n))\nonumber\\
&=& (1+o(1)){\rm vol}(G),
\eear
where the inequality follows from ${\rm vol}(S_i)$ will increase by a factor at least $\tilde{d}/2$ in each step as shown in \rlem{PLG-lower}. Similarly, we have ${\rm vol}_2(T_i) = (1+o(1)){\rm vol}_2(G).$

Next, we give some properties of $S_{i+1}$. Our goal is to find the expected weight of $S_{i+1}$, $\mathbb{E}[Y_i]$, and the variable $\nu_i$ (defined below) to apply \rlem{bino1}. To do this, define $X_j$ as the indicator function that node $j$ is in $S_{i+1}$. Thus, by union bound and a fact (in the proof of Lemma 3.2 in \cite{CL02}), we have 
\bear{xxx6}
{\rm vol}(S_i)w_j/{\rm vol}(G)-({\rm vol}(S_i)w_j/{\rm vol}(G))^2 \nonumber\\
\le \mathbb{P}(X_j = 1) \le {\rm vol}(S_i)w_j/{\rm vol}(G).
\eear
Let $Y_i$ be the volume of $S_{i+1}$, i.e., $Y_i = \sum_{j\in T_i}w_jX_j$. Thus, we have 
\beq{xxx7}
\mathbb{E}[Y_i] = (1+o(1)){\rm vol}(S_i){\rm vol}_2(T_i)/{\rm vol}(G).
\eeq
Similarly, define $\nu_i = \sum_{j\in T_i}w_j^2\mathbb{P}(X_j = 1)$. We have 
\beq{xxx8}
\nu_i = (1+o(1)){\rm vol}(S_i){\rm vol}_3(T_i)/{\rm vol}(G).
\eeq
Using the properties of $T_i$ and $S_{i+1}$, we now show the inductive step, namely: ${\rm vol}(S_{i+1}) \le (1+\delta)\tilde{d}{\rm vol}(S_i)$ with probability $1-o(\log^{-2}(n))$. 
Recalling \rlem{bino1}, we have 
\beq{xxx9}
\mathbb{P}(Y_i>\mathbb{E}[Y_i]+\kappa_i)\le \exp\left(-\frac{\kappa_i^2}{2(\nu_i+\tilde{m}_i\kappa_i/3)}\right).
\eeq
We need to consider three cases which are $3<\beta<4$, $\beta = 4$, and $\beta >4$. In each case, we first estimate $\nu_i$ and then compare $c\tilde{m}_i$ and $\sqrt{c\nu_i}$. According to the above comparison, we specify $\kappa_i$ for each case and conclude our desired result. Let $c = 10\log\log(n)$ and consider the three cases.

\noindent\textbf{Case 1 ($3<\beta < 4$):} First note that 
\bear{yyy1}
\nu_i &\approx& {\rm vol}(S_i)\int_{m}^{\tilde{m}_i}x^{3-\beta}ndx/C_1{\rm vol}(G) \nonumber\\
&=& \mathcal{O}({\rm vol}(S_i)\tilde{m}_i^{4-\beta}\log^{\beta-2}(n)).
\eear
Therefore, $c\tilde{m}_i \gg \sqrt{c\nu_i}$ for $n$ sufficiently large. Hence, choosing $\kappa_i = c\tilde{m}_i$, we have
\beq{yyy2}
\mathbb{P}(Y_i>\mathbb{E}[Y_i]+\kappa_i) \le \exp\left(-\frac{\kappa_i^2}{4\tilde{m}_i\kappa_i}\right) = o(\log^{-2}(n)).
\eeq
Note that by \req{xxx7}, we need only to show that $\kappa_i \le \delta {\rm vol}(S_i){\rm vol}_2(T_i)/{\rm vol}(G)$. This suffices to show 
\beq{yyy3}
{\rm vol}(S_i)\ge \left(\frac{c{\rm vol}(G)}{\delta {\rm vol}_2(T_i)}\right)^{\beta-2/\beta-3}\log^{\beta/\beta-3}(n).
\eeq
But this is true since the initial weight is greater than $\log^{\beta/\beta-3}(n)$ and it increases by a factor of at least $\tilde{d}/2$ in each step.

\noindent\textbf{Case 2 ($\beta = 4$):} We have $\nu_i = \mathcal{O}({\rm vol}(S_i)\log(\tilde{m}_i)\log^2(n))$ and $c\tilde{m}_i \gg \sqrt{c\nu_i}$. Hence, with similar computation as that described in Case 1, we have the desired result.

\noindent\textbf{Case 3 ($\beta > 4$):} We have $\nu_i = \Theta({\rm vol}(S_i)\log^2(n))$. By direct computation, we have $\sqrt{c\nu_i} \gg c\tilde{m}_i$ provided the initial weight is greater than $\log^{5/\beta-4}(n)$. Hence, we choose $\kappa_i = \sqrt{c\nu_i}$. Similarly, we just need to show that $\kappa_i \le \delta {\rm vol}(S_i){\rm vol}_2(T_i)/{\rm vol}(G)$. This is true since, by \req{xxx8},
\beq{yyy4}
{\rm vol}(S_i) \ge \frac{c{\rm vol}_3(T_i){\rm vol}(G)}{\delta^2{\rm vol}_2(T_i)^2} = \Theta(\log\log(n)).
\eeq

Note that the probability of failure in each step is $\mathcal{O}(\log^{-2}(n))$ and there are at most $\epsilon\log_{\tilde{d}}(n)+\lambda$ steps. Thus, the sum of weights of nodes in $\epsilon\log_{\tilde{d}}(n)+\lambda$-neighborhood of $x$ is at most $2W\tilde{d}^\lambda n^{\epsilon(1+o(1))}$. We conclude that the desired result holds with probability $1-\mathcal{O}(\log^{-1}(n))$ as each node has weight at least $\log(n)$.
\eproof

\bsec{Appendix B}{appenb}
\subsection{Proofs of lemmas in \rsec{lem}}
In this section, we first show \rlem{dist}, \rlem{rec}, and \rlem{node}. The proofs of these lemmas use techniques for balls and bins problems. To solve a problem like these, in general, we first find a proper target function and write the target function as a sum of indicator functions. We next use a large deviation result, e.g. \rlem{bino1}, to show the target function is concentrated around its mean. The proofs of the three lemmas do follow the above procedure and are shown below.

\bproof (\rlem{dist})
Let $X_i$ be the indicator function that the distance between the first node and node $i$ is smaller than $\sqrt{64n\log(n)/\pi k}$. Let $Y = \sum_{i=2}^{k+1}X_i$, $i.e.$, $Y$ is the number of nodes with distance to the first node smaller than $\sqrt{64n\log(n)/\pi k}$. To show this lemma, we first find out the mean of $Y$ and show $Y$ is around $\mathbb{E}[Y]$ with overwhelming probability. Indeed,
\bear{app0}
\mathbb{E}[Y] &=& \sum_{i=2}^{k+1}\mathbb{E}[X_i] \nonumber \\
&=& \sum_{i=2}^{k+1}\mathbb{P}(X_i = 1) \nonumber \\
&\ge& 16\log(n)
\eear
where the last inequality follows if the first node is located at a corner of the square.

To apply \rlem{bino1}, we choose $c = \mathbb{E}[Y]/2$ and observe $\nu = \mathbb{E}[Y]$, and get
\beq{app1}
\mathbb{P}(Y \le \mathbb{E}[Y]/2) \le \exp(-(\mathbb{E}[Y]/2)^2/2\mathbb{E}[Y]) \le n^{-2}
\eeq
The above equation \req{app1} implies at least $8\log(n)$ nodes close to the first node with probability $1-n^{-2}$ and we have the lemma.
\eproof

Similar to the above proof, we show the rest of two lemmas.

\bproof (\rlem{rec})
We may assume $A \ge 10\log(n)$. Let $Y_i$ be the indicator function that node $i$ falls in that rectangle. Let $X = \sum_{i=1}^{n}Y_i$, $i.e.$, $X$ is the number of nodes falling in the rectangle. To show this lemma, we first find out the mean of $X$ and show $X$ is around $\mathbb{E}[X]$ with overwhelming probability. Indeed,
\bear{app2}
\mathbb{E}[X] &=& \sum_{i=1}^{n}\mathbb{E}[Y_i] \nonumber \\
&=& \sum_{i=1}^{n}\mathbb{P}(Y_i = 1) \nonumber \\
&\ge& 10\log(n).
\eear

To apply \rlem{bino1}, we choose $c = \mathbb{E}[X]$ and observe $\nu = \mathbb{E}[X]$, and get
\beq{app3}
\mathbb{P}(X \ge 2\mathbb{E}[X]) \le \exp(-(\mathbb{E}[X])^2/8\mathbb{E}[X]/3) \le n^{-2}
\eeq
The above equation \req{app3} implies at most $2A$ nodes falling in the rectangle with probability $1-n^{-2}$ and we have the lemma.
\eproof

\bproof (\rlem{node})
Let $Y_j$ be the indicator function that $(i, j) \in E$. Let $X = \sum_{j\neq i}Y_j$, $i.e.$, $X$ is the number of one-hop neighbors of node $i$. To show this lemma, we first find out the mean of $X$ and show $X$ is around $\mathbb{E}[X]$ with overwhelming probability. Indeed,
\bear{app5}
\mathbb{E}[X] &=& \sum_{j\neq i}\mathbb{E}[Y_j] \nonumber \\
&=& (1+o(1))w_i.
\eear

To apply \rlem{bino1}, we choose $c_1 = \mathbb{E}[X]$ for upper bound and $c_2 = \mathbb{E}[X]/2$ for lower bound. Observe $\nu = \mathbb{E}[X]$, and get
\beq{app6}
\mathbb{P}(X \ge 2\mathbb{E}[X]) \le \exp(-(\mathbb{E}[X])^2/8\mathbb{E}[X]/3) \le o(n^{-1})
\eeq
\beq{app7}
\mathbb{P}(X \le \mathbb{E}[X]/2) \le \exp(-(\mathbb{E}[X]/2)^2/2\mathbb{E}[X]) \le o(n^{-1})
\eeq

One may observe $o(1)$ term does not affect the results. Therefore, with above equations \req{app6} and \req{app7}, we have the lemma.
\eproof

\subsection{Proofs of \rthe{flooding} and \rthe{PLG-lower1}}
In this section, we present our proofs for \rthe{flooding} and \rthe{PLG-lower1}, the performance of our algorithm and the lower bound on the file dissemination time of any possible algorithm. We consider a random power law graph with $\beta>2$ and nodes are only allowed to download the file from nodes at most two hops away. In \rthe{flooding}, we set the input of Algorithm 1 as $\epsilon = 0$ and $\mathcal{L} = \infty$. Thus, nodes always request to one-hop neighbors on the social-graph. We show that our load-balancing scheme, exploiting the property social networks have small diameters, guarantees the file dissemination time scales like $\sqrt{n}$. In the proof of \rthe{PLG-lower1}, we adopt an approach similar to that in \rthe{PLG-lower2} in which we find a lower bound on the transport load. We show that the performance of our algorithm only differs from the best possible file dissemination time by a factor of $n^\xi$ for any $\xi>0$. Our proofs are presented in the follows.

\bproof (\rthe{flooding})
We first claim that each transmission has a rate $\Omega(1/\sqrt{n})$. To show this, consider a horizontal highway node and its corresponding stripe. Note that these $n$ nodes are placed uniformly and independently on the square. By \rlem{rec}, there are $\mathcal{O}(\sqrt{n})$ nodes in this stripe with high probability. On the other hand, each node only generates at most 6 flows. Therefore, each flow through the horizontal highway node can have a rate of $\Omega(1/\sqrt{n})$. A similar argument applies to vertical highway nodes. As this is true for all highway nodes with high probability, we have the claim. In addition, there exists a constant $c_1$ such that each node can receive the file in $c_1\sqrt{n}F$ time slots since the transmission starts.

Similar to the proof in \rthe{neighborsearch}, we show the theorem by induction on $k$: the distance from a node to the source on the social-graph. Our claim is nodes at distance $k$ to the source can receive the file in $c_1k\sqrt{n}\log_2(n)F$ time slots. It is clear that the base case is true for $k=1$. Suppose this is true for $k-1$ and consider nodes at distance $k$ to the source. Since each such node is not inactive at time $c_1(k-1)\sqrt{n}\log_2(n)F$, the node must be in a binary true with an active node as the root. Therefore, this node has to wait at most $\log_2(n)-1$ transmissions before getting served. Thus, the node can receive the file at time $c_1k\sqrt{n}\log_2(n)F$. Hence, by mathematical induction, we have the claim. Note that the diameter of the social graph is $\mathcal{O}(\log(n))$. All nodes can get the file in $\mathcal{O}(\sqrt{n}\log^2(n)F)$ time slots.
\eproof

\bproof (\rthe{PLG-lower1})
To apply \rlem{bmlimit}, we just need to show that the transport load is $\Omega(n^{3/2-\xi}F)$ for any $\xi>0$ with probability $1-o(1)$. The idea is to show there are $\Theta(n)$ nodes in the largest component which only have small-sized 2-neighborhoods. Thus, these nodes must download the file from nodes which are geographically far away from them on the wireless-square. We first state the flow of the proof. In the first step, we claim we only need to consider a random power law graph with minimum expected degree $m = K\log(n)$ for some $K\ge 10$. More precisely, only consider nodes with weight in the region $[K\log(n)$  $2K\log(n)]$ in such graphs. We next show that only a vanishing fraction of the number of them can find geographic proximate one-hop or two-hop neighbors. In the end, we show that $\Theta(n)$ of them are indeed in the largest component and thus, have the theorem.

We first show that we may assume that the minimum expected degree $m = K\log(n)$ for some constant $K \ge 10$. To do this, consider the original minimum expected degree $\hat{m} < 10\log(n)$ and the original expected degree sequence $\hat{w} = (\hat{w}_1, \ldots, \hat{w}_n)$. Let $w$ be the expected degree sequence for $m = K\log(n)$ for some $K\ge 10$. Observe that \req{weight1} is an increasing function in terms of $\bar{d}$. We have $\hat{w}$ is smaller than $w$ term by term. Thus, by coupling, the random power law graph generated by $w$ contains the original random power law graph stochastically.

Next, we show that nodes with weight in that region have a small-sized 2-neighborhood with high probability. This property is important as small-sized neighborhood implies it is hard to find geographic proximate neighbors. Consider $2\xi>\eta>0$. Let $\mathcal{N}_i$ be the set of nodes that node $i$ can reach in 2 hops in the social-graph. We claim $\mathbb{P}(|\mathcal{N}_i| \le 10Kn^\eta\log(n)) = 1-o(1/\log(n))$. Indeed, by \rlem{node}, node $i$ has at most $4K\log(n)$ neighbors on the social-graph with probability $1-o(n^{-1})$. Further, the probability that one of its neighbors is of weight greater than $n^\eta$ is smaller than
\beq{lower1}
\frac{2K\log(n)\int_{n^\eta}^Mx^{1-\beta}ndx}{{\rm vol(G)}\int_{K\log(n)}^Mx^{-\beta}dx} = o(1/\log(n)).
\eeq
Thus, the sum of weights of its neighbors is smaller than $4Kn^\eta\log(n)$ with probability $1-o(1/\log(n))$. Hence, by \rlem{node} again, we have the claim.

Let $\mathcal{M}$ be the set of nodes with expected degree in the range $[K\log(n)$ $2K\log(n)]$. Then, 
\beq{lower2}
|\mathcal{M}| \approx \frac{\int_{K\log(n)}^{2K\log(n)}x^{-\beta}ndx}{\int_{K\log(n)}^Mx^{-\beta}dx} = (1+o(1))(1-2^{1-\beta})n
\eeq

We next claim only $o(n)$ nodes in $\mathcal{M}$ have geographic proximate neighbors (the distance between the neighbors and the node is smaller than $\sqrt{n/10K\pi n^\eta\log^2(n)}$). This property along with \req{lower1} implies almost all nodes in $\mathcal{M}$ do not have geographic proximate neighbors. We show this property in the follows. We first find out the expected number of nodes in $\mathcal{M}$ which have geographic proximate neighbors. Let $X_i$ be the indicator function that the Euclidean distance from node $i$ to $\mathcal{N}_i$ on the wireless-square is smaller than $\sqrt{n/10K\pi n^\eta\log^2(n)}$. Therefore, we have, for $i \in \mathcal{M}$ and $n$ large enough
\beq{lower3}
\mathbb{P}(X_i = 1) \le 1/\log(n) + o(1/\log(n)) \le 2/\log(n)
\eeq
since the first term is the probability that $|\mathcal{N}_i| \le 10Kn^\eta\log(n)$ and $X_i = 1$, and the second term is the probability that $|\mathcal{N}_i| > 10Kn^\eta\log(n)$. Therefore, we have
\beq{lower4}
\mathbb{E}\left[\sum_{i\in \mathcal{M}}X_i\right] \le 2(1-2^{1-\beta})n/\log(n)
\eeq

Next, we show a concentration result. We claim that $\mathbb{P}(|\sum_{i\in \mathcal{M}}X_i-\mathbb{E}[\sum_{i\in \mathcal{M}}X_i]|\ge n/\log^{1/3}(n)) = o(1).$ Indeed, by Chebyshev's inequality, we have
\bear{lower5}
&&\mathbb{P}\left(\left|\sum_{i\in \mathcal{M}}X_i-\mathbb{E}\left[\sum_{i\in \mathcal{M}}X_i\right]\right|\ge n/\log^{1/3}(n)\right)\nonumber\\
&&\le \frac{\mathbb{E}[(\sum_{i\in \mathcal{M}}X_i)^2]}{n^2/\log^{2/3}(n)}\nonumber\\
&&\le \frac{2n^2/\log(n)}{n^2/\log^{2/3}(n)} = o(1)
\eear
where the last inequality follows from $\mathbb{E}[X_iX_j] \le \mathbb{E}[X_i]$.

At the end, let $\mathcal{S}$ be the set of nodes in the largest component. Since almost all nodes are in the largest component, we have $|\mathcal{S}\cap \mathcal{M}| = \Theta(n).$ Hence, the result follows by the above fact only $o(n)$ nodes in $\mathcal{M}$ have geographic proximate one-hop or two-hop neighbors.
\eproof

\end{document}